\documentclass[conference,a4paper]{APSIPA2021}
\usepackage{multirow}
\usepackage[dvips]{graphicx}
\usepackage{amsmath, bm, physics}
\usepackage[psamsfonts]{amssymb}
\usepackage{amsxtra}
\usepackage{threeparttable}
\usepackage{cite}
\usepackage{citesort}
\usepackage{siunitx}

\newcommand{\scm}[2]{R_{ #1 }^{(\mathrm{ #2 })}}
\newcommand{\scmd}[2]{ { {R'} _ { #1 } ^ { (\mathrm{ #2 }) } } }
\newcommand{\covar}[2]{r_{ #1 }^{(\mathrm{ #2 })}}

\begin{document}
\bstctlcite{IEEEexample:BSTcontrol}
\title{Speech Enhancement by Noise Self-Supervised Rank-Constrained Spatial Covariance Matrix Estimation via Independent Deeply Learned Matrix Analysis}

\author{%
\authorblockN{%
Sota~Misawa\authorrefmark{1},
Norihiro~Takamune\authorrefmark{1},
Tomohiko~Nakamura\authorrefmark{1},
Daichi~Kitamura\authorrefmark{2},
Hiroshi~Saruwatari\authorrefmark{1},\\
Masakazu~Une\authorrefmark{3}
and
Shoji~Makino\authorrefmark{3}\authorrefmark{4}
}
\authorblockA{%
\authorrefmark{1}
The University of Tokyo, Tokyo, Japan \\}
\authorblockA{%
\authorrefmark{2}
National Institute of Technology, Kagawa College, Kagawa, Japan}
\authorblockA{%
\authorrefmark{3}
University of Tsukuba, Ibaraki, Japan}
\authorblockA{%
\authorrefmark{4}
Waseda University, Fukuoka, Japan}
}

\maketitle
\thispagestyle{empty}

\begin{abstract}
Rank-constrained spatial covariance matrix estimation (RCSCME) is a method for the situation that the directional target speech and the diffuse noise are mixed. 
In conventional RCSCME, independent low-rank matrix analysis (ILRMA) is used as the preprocessing method.
We propose RCSCME using independent deeply learned matrix analysis (IDLMA), which is a supervised extension of ILRMA.
In this method, IDLMA requires deep neural networks (DNNs) to separate the target speech and the noise.
We use Denoiser, which is a single-channel speech enhancement DNN, in IDLMA to estimate not only the target speech but also the noise.
We also propose noise self-supervised RCSCME, in which  we estimate the noise-only time intervals using the output of Denoiser and design the prior distribution of the noise spatial covariance matrix for RCSCME. 
We confirm that the proposed methods outperform the conventional methods under several noise conditions.
\end{abstract}

\section{Introduction}
In this study, we deal with speech enhancement. In particular, we deal with 
the problem of extracting directional target speech from diffuse noise.
This situation arises, for example, when the speaker is close to the microphones in a noisy environment such as a crowded place or a train station. 

From a practical perspective, it is effective to use blind source separation (BSS) \cite{sawada2019review}, which does not require any spatial information or characteristics of each source.
In a determined or overdetermined situation (number of microphones $M$ $\geq$ number of sources $N$), independent vector analysis (IVA) \cite{hiroe2006solution,kim2006blind,ono2011stable} and independent low-rank matrix analysis (ILRMA) \cite{kitamura2016determined} have been proposed for BSS. ILRMA assumes that the power spectrogram for each source is modeled by nonnegative matrix factorization (NMF) \cite{lee1999learning}. Since these methods assume that the rank of the spatial covariance matrix (SCM) of each source is unity,
we call these methods \textit{rank-1 methods}.
In a rank-1 method, the noise remains in the separated target signal when the noise has a full-rank SCM, such as diffuse noise \cite{published_papers/7980945,araki2003equivalence,published_papers/7980919}.

Multichannel NMF (MNMF) \cite{ozerov2009multichannel,sawada2013multichannel} and FastMNMF \cite{ito2019fastmnmf,sekiguchi2019fast} have been proposed as methods to model full-rank SCMs for each source. MNMF adopts the full-rank nature for the source model \cite{duong2010under} and can handle diffuse sources, but it is computationally expensive and highly dependent on initial values \cite{kitamura2016determined}. In FastMNMF, the computational complexity is reduced by assuming that the SCMs are jointly diagonalizable. However, the problem that the performance is not robust to changes in the initial value still remains.

Rank-constrained spatial covariance matrix estimation (RCSCME) \cite{kubo2019efficient,kubo2020blind} has been proposed as a method of blind speech enhancement for mixed signals of directional target speech and diffuse noise. RCSCME uses a rank-1 method such as ILRMA in advance. Since a rank-1 method accurately obtains $M-1$ noise signals, we can obtain the rank-($M-1$) component of the noise SCM from these signals. RCSCME estimates the time-varying variance of the target speech and the diffuse noise, and complements the deficient rank-1 component of the noise SCM at the same time. As shown in \cite{kubo2019efficient}, this method is robust and converges rapidly because there are fewer estimated parameters than in MNMF and FastMNMF.

For the case that training data are available, many supervised source separation methods using a deep neural network (DNN) have been proposed. For a determined or overdetermined situation, independent deeply learned matrix analysis (IDLMA) \cite{makishima2019independent} has been proposed to improve source separation performance by changing the source model of ILRMA from NMF to DNNs. Owing to the appropriately trained DNNs, IDLMA achieves higher performance than ILRMA.
Since IDLMA is a rank-1 method similarly to ILRMA, the noise remains in the separated target signal in a situation with diffuse noise.

In this paper, we propose the use of IDLMA, which is a rank-1 method, as the preprocessing method of RCSCME.
Since IDLMA was originally proposed for musical source separation, we should prepare DNNs that estimate the power spectrogram of speech or noise.
One of the DNN-based single-channel speech enhancement methods is Denoiser~\cite{defossez2020real}, which estimates the target speech from a mixed-signal input in the waveform domain. 
Whereas IDLMA requires DNNs for each source, Denoiser extracts only the target speech. Thus, we also propose a scheme to estimate the power spectrogram of the noise using Denoiser. 
Furthermore, considering the high capability of Denoiser to reduce noise, we can find the noise-only time intervals of the observed signal using the output of Denoiser. 
The SCM of this noise-only signal is a good approximation of the noise SCM.
Thus, we design the prior distribution of the noise using this approximated noise SCM for RCSCME, and we call this method \textit{noise self-supervised RCSCME}. 
We derive the new update rule on the basis of the EM algorithm for noise self-supervised RCSCME. We conduct a simulated experiment to verify the effectiveness of the proposed methods.

\section{Conventional Methods}
\subsection{ILRMA and IDLMA}
Let $\bm{s}_{ij}=(s_{ij,1},\ldots,s_{ij,N})^\mathsf{T}$, $\bm{x}_{ij}=(x_{ij,1},\ldots,x_{ij,M})^\mathsf{T}$, and $\bm{y}_{ij}=(y_{ij,1},\ldots,y_{ij,N})^\mathsf{T}$
denote the short-time Fourier transform (STFT) of the source, observed, and separated signals, respectively, where $i=1,\ldots, I$ and $j=1,\ldots, J$ are the indices of the frequency bin and time frame, respectively. Here, ${}^\mathsf{T}$ denotes transpose. We also represent the spectrograms for each signal as $S_n, X_m, Y_n\in \mathbb{C}^{I\times J}$ whose $(i,j)$th elements are $s_{ij, n}, x_{ij, m}$, and $y_{ij, n}$, where $n=1,\ldots, N$ and $m=1,\ldots, M$ are the source and channel indices, respectively.

When the reverberation time is sufficiently shorter than the window length of the STFT and each source is a point source,
we can denote the observed signal as
\begin{align}
    \bm{x}_{ij} = A_i\bm{s}_{ij},
\end{align}
where $A_i=\qty(\bm{a}_{i, 1}, \ldots ,\bm{a}_{i, N})\in \mathbb{C}^{M\times N}$ is a mixing matrix.
Assuming that $N=M$ and $A_i$ is invertible, we can obtain the separated signal by estimating its inverse matrix $W_i = \qty(\bm{w}_{i, 1}, \ldots ,\bm{w}_{i, N})^\mathsf{H}\in\mathbb{C}^{N\times M}$, which is called the demixing matrix, such that
\begin{align}
    \bm{y}_{ij} = W_i\bm{x}_{ij},
\end{align}
where ${}^\mathsf{H}$ denotes the Hermitian transpose.

In ILRMA\cite{kitamura2016determined} and IDLMA\cite{makishima2019independent}, 
it is assumed that the separated signals are mutually independent and that the $n$th separated signal in each time-frequency frame is generated by the univariate complex Gaussian distribution as
\begin{align}
    y_{ij,n}\sim \mathcal{N}_\mathrm{c}\left(0, \sigma_{ij,n}^2\right),
\end{align}
where $0$ and $\sigma_{ij,n}^2$ are the mean and variance, respectively. We use the negative log-likelihood of the observed signal as the cost function:
\begin{align}
    \mathcal{L}_{\mathrm{I}} =& - \log \prod_{i,j,n}p(x_{ij,n})\nonumber\\
    =& - \log \prod_{i,j,n}p(y_{ij,n}) - 2J\sum_i \log |\det W_i|\nonumber\\
    =& \sum_{i,j,n}\qty(\frac{|w_{in}^\mathsf{H}x_{ij}|^2}{\sigma_{ij,n}^2} + 2\log \sigma_{ij,n})\nonumber\\
    & -2J\sum_i \log |\det W_i| + \mathrm{const.},
\end{align}
where $\mathrm{const.}$ includes the terms independent of the target variables.
In ILRMA, $\sigma_{ij,n}^2$ is modeled by NMF \cite{lee1999learning} as
\begin{align}
    \sigma_{ij,n}^2 = \sum_k t_{ik, n}v_{kj, n},
\end{align}
where $t_{ik, n}\geq 0$ and $v_{kj, n}\geq 0$ are the NMF variables. 
Instead, in IDLMA, we use a source separation DNN that estimates the power spectrogram:
\begin{align}
    \sigma_{ij,n}^2 &= \max \qty{\abs{\qty(\mathrm{DNN}_n \qty(Y_n))_{ij}}^2, \varepsilon},\label{eq:epsilon1}
\end{align}
where $\varepsilon>0$ is a small number used to avoid numerical instability.
Here, $\mathrm{DNN}_n$ is the trained DNN corresponding to the $n$th source and its input is the current separated signal.

Since diffuse noise is not a point source, ILRMA and IDLMA cannot separate noise in the same direction as the directional target speech in principle \cite{published_papers/7980945,araki2003equivalence,published_papers/7980919}.
Therefore, we cannot directly apply ILRMA and IDLMA to diffuse noise reduction, which often arises in practical speech enhancement application.

\subsection{RCSCME}

RCSCME \cite{kubo2019efficient,kubo2020blind} deals with the problem of speech enhancement when directional target speech and diffuse noise are mixed. 
In \cite{kubo2019efficient}, it is assumed that the generative model of the observed signal is the multivariate complex Gaussian distribution 
\begin{align}
    \label{eq:xdist}
    p(x_{ij};0, \scm{ij}{o}) \propto \frac{\exp(-x_{ij}^\mathsf{H}\scm{ij}{o}x_{ij})}{\det \scm{ij}{o}},
\end{align}
where $\scm{ij}{o}$ denotes the covariance matrix of the observed signal. $\scm{ij}{o}$ is modeled as
\begin{align}
    \scm{ij}{o} = \covar{ij}{t}\scm{i}{t} + \covar{ij}{n}\scm{i}{n},
\end{align}
where $\covar{ij}{t}>0$ and $\covar{ij}{n}>0$ are the variances of the target speech and the noise, and $\scm{i}{t}$ and $\scm{i}{n}$ are the SCMs of the target speech and the noise, respectively. 
To induce the sparsity of the target speech, we assume the inverse gamma distribution for the prior distribution on $\covar{ij}{t}$,
\begin{align}
\label{eq:rprior}
    p(\covar{ij}{t}; \alpha, \beta)\propto
    \qty(\covar{ij}{t})^{-\alpha -1}\exp(-\frac{\beta}{\covar{ij}{t}}),
\end{align}
where $\alpha>0$ and $\beta >0$ are the shape parameter and scale parameter, respectively.
Using the estimated mixing matrix $W_i^{-1}=A_i$, we model the SCMs of the target source and noise as
\begin{align}
    \scm{i}{t} &= \bm{a}_{i,n_t}\bm{a}_{i,n_t}^\mathsf{H},\\
    \scm{i}{n} &= \scmd{i}{n} + \lambda_i \bm{v}_i\bm{v}_i^\mathsf{H},\\
    \scmd{i}{n}&=\frac{1}{J}\sum_j \hat{y}_{ij}^{(\mathrm{n})} \qty(\hat{y}_{ij}^{(\mathrm{n})})^\mathsf{H},\label{eq:noisescm}\\
    \hat{y}_{ij}^{(\mathrm{n})} &= A_i
    \left(
        w_{i,1}^\mathsf{H}x_{ij},\ldots , w_{i,n_t-1}^\mathsf{H}x_{ij}, 0,\nonumber \right. \\
    & \qquad \qquad \quad \left.
        w_{i,n_t+1}^\mathsf{H}x_{ij}, \ldots , w_{i,N}x_{ij}
    \right)^\mathsf{T},
\end{align}
where $n_t$ denotes the target index and $\scmd{i}{n}\in \mathbb{C}^{M\times M}$ is the rank-($M-1$) SCM of the diffuse noise estimated by a rank-1 method. Since the rank of $\scmd{i}{n}$ is $M-1$, $\bm{v}_i$ must be linearly independent of any column vectors of $\scmd{i}{n}$ in order that $\scm{i}{n}$ has full rank. For example, the unit eigenvector corresponding to the zero eigenvalue of $\scmd{i}{n}$ satisfies this condition for $\bm{v}_i$. RCSCME simultaneously estimates $\lambda_i$, $\covar{ij}{t}$, and $\covar{ij}{n}$ by maximum a posteriori estimation. 
Using (\ref{eq:xdist}) and (\ref{eq:rprior}), we calculate the log-posterior of the observed signal as
\begin{align}
    \mathcal{L}_{\mathrm{RC}}(\Theta_{\mathrm{c}}) =& \log \prod_{i,j} p(\bm{x}_{ij}|\Theta_{\mathrm{c}})p(\covar{ij}{t}; \alpha, \beta)\nonumber \\
    =& -\sum_{i,j}\Biggl(\bm{x}_{ij}^\mathsf{H}\scm{ij}{o}\bm{x}_{ij} + \log \det \scm{ij}{o}\Biggr.\nonumber \\
    & + \left.(\alpha + 1)\log \covar{ij}{t} + \frac{\beta}{\covar{ij}{t}}\right) + \mathrm{const.},
\end{align}
where $\Theta_{\mathrm{c}} = \{\lambda_i, \covar{ij}{t}, \covar{ij}{n}\}$. We maximize the log-posterior function using the EM algorithm with the following $Q$ function (see \cite{kubo2019efficient} for details):
\begin{align}
    Q_{\mathrm{RC}}\qty(\Theta_{\mathrm{c}}; \tilde{\Theta}_{\mathrm{c}})=& -\sum_{i,j}\Bigg((\alpha + 2)\log \covar{ij}{t} +\frac{\hat{r}_{ij}^{(\mathrm{t})} + \beta}{\covar{ij}{t}}\nonumber \\
    & +M\log \covar{ij}{n} +\log \det \scm{i}{n}\nonumber\\
    &+ \frac{\tr(\hat{R}_{ij}^{(\mathrm{n})}\qty(R_{i}^{(\mathrm{n})})^{-1})}{\covar{ij}{n}}\Bigg)+ \mathrm{const.},
\end{align}
where $\tilde{\Theta}_{\mathrm{c}} = \{\tilde{\lambda}_i, \tilde{r}_{ij}^{(\mathrm{t})}, \tilde{r}_{ij}^{(\mathrm{n})}\}$ is the set of the up-to-date parameters, and $\hat{r}_{ij}^{(\mathrm{t})}$ and $\hat{R}_{ij}^{(\mathrm{n})}$ are calculated in the E-step as follows:
\begin{align}
    \tilde{R}_{i}^{(\mathrm{n})}=&\scmd{i}{n} + \tilde{\lambda}_i\bm{v}_i\bm{v}_i^\mathsf{H},\\
    \scm{ij}{o} =& \tilde{r}_{ij}^{(\mathrm{t})}\scm{i}{t} + \tilde{r}_{ij}^{(\mathrm{n})}\tilde{R}_{i}^{(\mathrm{n})},\\
    \hat{r}_{ij}^{(\mathrm{t})}=&\tilde{r}_{ij}^{(\mathrm{t})}-\qty(\tilde{r}_{ij}^{(\mathrm{t})})^2\bm{a}_{i,n_t}^\mathsf{H}\qty(\scm{ij}{o})^{-1}\bm{a}_{i,n_t}\nonumber \\
    &+ \qty|\tilde{r}_{ij}^{(\mathrm{t})}\bm{x}_{ij}^\mathsf{H}\qty(\scm{ij}{o})^{-1}\bm{a}_{i,n_t}|^2,\\
    \hat{R}_{ij}^{(\mathrm{n})} =& 
    \tilde{r}_{ij}^{(\mathrm{n})}\tilde{R}_{i}^{(\mathrm{n})} - \qty(\tilde{r}_{ij}^{(\mathrm{t})})^2
    \tilde{R}_{i}^{(\mathrm{n})} \qty(\scm{ij}{o})^{-1}\tilde{R}_{i}^{(\mathrm{n})}\nonumber\\
    &+ \qty(\tilde{r}_{ij}^{(\mathrm{t})})^2
    \tilde{R}_{i}^{(\mathrm{n})} \qty(\scm{ij}{o})^{-1}
    \bm{x}_{ij}\bm{x}_{ij}^{\mathsf{H}}
    \qty(\scm{ij}{o})^{-1}\tilde{R}_{i}^{(\mathrm{n})}.
\end{align}
In the M-step, by finding the stationary point of $Q_{\mathrm{RC}}$, we obtain the update rules as 
\begin{align}
    \covar{ij}{t}&\leftarrow \frac{\hat{r}_{ij}^{(\mathrm{t})}+\beta}{\alpha + 2},\label{eq:targetcovarruleold}\\
    \lambda_i &\leftarrow \bm{u}_i^\mathsf{H}\qty(\frac{1}{J}\sum_j \frac{1}{\tilde{r}_{ij}^{(\mathrm{n})}}\hat{R}_{ij}^{(\mathrm{n})})\bm{u}_i,\label{eq:lambdaruleold}\\
    \scm{i}{n} &\leftarrow \scmd{i}{n} + {\lambda}_i\bm{v}_i\bm{v}_i^\mathsf{H},\\
    \covar{ij}{n}&\leftarrow \frac{1}{M}\tr(\hat{R}_{ij}^{(\mathrm{n})}\qty(R_{i}^{(\mathrm{n})})^{-1}),\label{eq:noisecovarruleold}
\end{align}
where $\bm{u}_i$ is the eigenvector corresponding to the zero eigenvalues of $\scmd{i}{n}$ satisfying $\bm{u}_i^{\mathsf{H}}\bm{v}_i=1$.
Here, we use the following property for $\scm{i}{n} = \scmd{i}{n} + \lambda_i \bm{v}_i\bm{v}_i^\mathsf{H}$, as proved in \cite{kubo2020blind}, in obtaining the update rule of $\lambda_i$:
\begin{align}
    \log \det \scm{i}{n}&= \log \lambda_i + \mathrm{const.},\label{eq:logdetR}\\
    \qty(\scm{i}{n})^{-1}&=\qty(E - \bm{u}_i\bm{v}_i^{\mathsf{H}})\qty(\scmd{i}{n})^{+}\qty(E - \bm{v}_i\bm{u}_i^{\mathsf{H}}) + \frac{1}{\lambda_i} \bm{u}_i\bm{u}_i^\mathsf{H},\label{eq:Rinverse}
\end{align}
where $E\in \mathbb{C}^{M\times M}$ is the identity matrix and ${}^{+}$ denotes the Moore--Penrose inverse.

Finally, we obtain the extracted target speech using the multichannel Wiener filter as 
\begin{align}
    \hat{s}_{ij}^{(\mathrm{t})} = \covar{ij}{t}\scm{i}{t}\qty(\scm{ij}{o})^{-1}\bm{x}_{ij}.
\end{align}

\section{Proposed Method}
\subsection{Motivation and Framework}

In recent years, research on DNNs has become more popular and DNN-based source separation methods have been proposed \cite{grais2014deep,defossez2020real,qian2018deep,nugraha2016multichannel,makishima2019independent}.
IDLMA \cite{makishima2019independent} has been proposed for determined audio source separation. IDLMA has higher separation performance than ILRMA by introducing the supervised DNNs into the source models. 
Since IDLMA is also a rank-1 method similarly to ILRMA, we propose the use of IDLMA instead of ILRMA as the preprocessing method for RCSCME.
We expect that RCSCME will achieve high separation performance using highly accurate parameters pre-estimated by IDLMA.
However, since the original IDLMA dealt with musical source separation, we should prepare new DNNs for the speech and the noise.
For the single-channel speech enhancement DNN, Denoiser \cite{defossez2020real} has been proposed. Here, we propose a scheme to estimate the power spectrogram not only of the target speech, but also of the noise using Denoiser.

As shown in Table \ref{tb:Denoiser}, 
Denoiser tends to have very high accuracy in terms of the source-to-inference ratio (SIR)~\cite{vincent2006performance} improvement, which is a metric of denoising, but poor results in terms of the sources-to-artifact ratio (SAR)~\cite{vincent2006performance}, which is a metric of artifacts. In other words, Denoiser is very good at removing noise, but it also tends to partially remove the target speech. 
Considering the high SIR of Denoiser, we can extract the noise-only time intervals from the observed signal corresponding to the quiet time intervals of the output of Denoiser and approximate the noise SCM from these intervals.
Thus, we propose noise self-supervised RCSCME, in which the prior distribution of the noise SCM is introduced using the approximated noise SCM.
\begin{table}[tb]
\caption{SIR improvement and SAR of Denoiser}
\label{tb:Denoiser}
\begin{center}
\begin{tabular}{c|cccc}
\hline              & \multicolumn{4}{c}{Noise}         \\ \cline{2-5}
              & Babble & Cafe  & Station & Traffic \\ \hline
SIR improvement {[}dB{]} & 17.45  & 19.23 & 18.33   & 20.64   \\
SAR {[}dB{]} & 8.89  & 9.94 & 7.54   & 10.83 \\ \hline 
\end{tabular}
\end{center}
\end{table}

\begin{figure*}[t]
\begin{center}
\includegraphics[width=180mm]{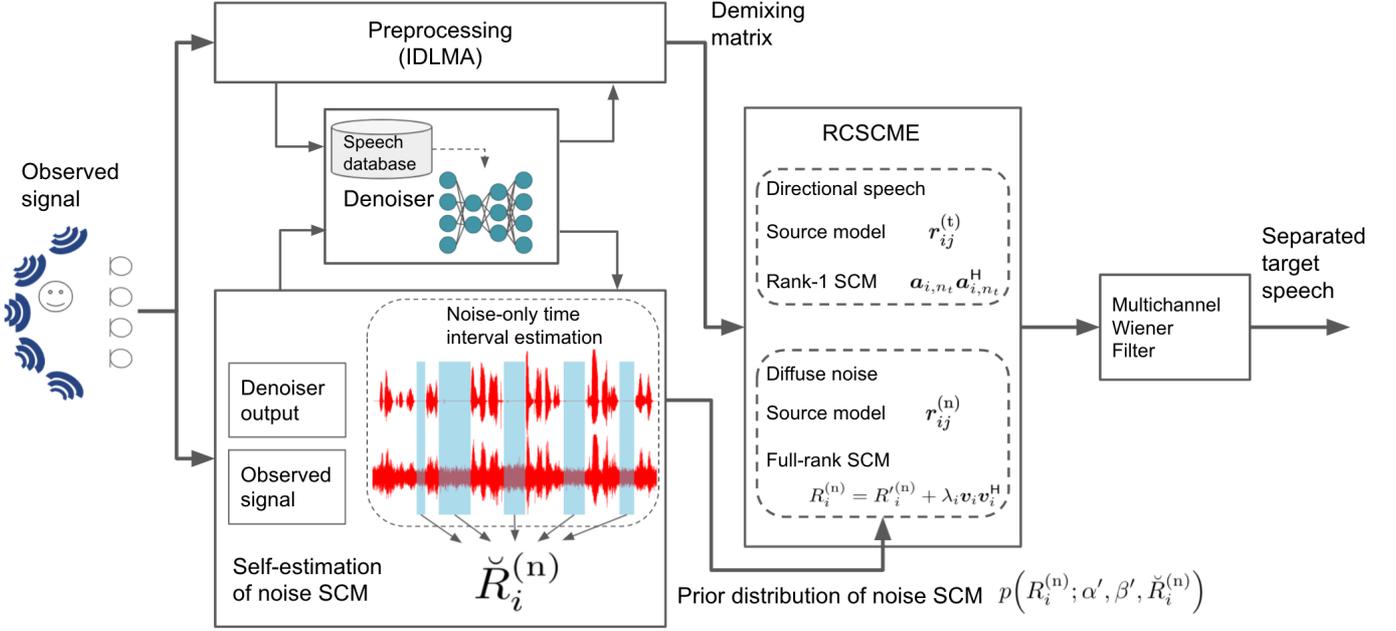}
\end{center}
\vspace*{-13pt}
\caption{Process flow of noise self-supervised RCSCME.}
\label{fig:flow}
\end{figure*}

Finally, an overview of the proposed method is as follows:
\begin{enumerate}
    \item Prepare the trained Denoiser. 
    \item Estimate the demixing matrix $W_i$ by IDLMA using Denoiser. 
    \item Calculate the quiet time intervals of the output using Denoiser, and set the corresponding time intervals of the observed signal as the noise-only signal. 
    \item Determine the prior distribution of the noise from the noise-only signal calculated in 3) and perform RCSCME with the result of 2).
    \item Construct the multichannel Wiener filter with the result of 4).
\end{enumerate}
The process flow of the proposed method is shown in Fig. \ref{fig:flow}. 

\subsection{IDLMA for Speech Enhancement}
In the separated signals $Y_n$ obtained by IDLMA, the $n_t$th signal must be the target speech, and the other channels must be noise signals. Therefore, for the variance of the $n_t$th separated signal, we use the output of the DNN, and for the other channels, we subtract the outputs of the DNN from the separated signals to simulate the estimation of the noise signal:
\begin{align}
    \zeta_{ij,n} &= \begin{cases}
        \qty(\mathrm{DNN}\qty(Y_n))_{ij}& \text{if } n=n_t,\\
        y_{ij,n} - \qty(\mathrm{DNN}\qty(Y_n))_{ij}& \text{otherwise,}
    \end{cases}\label{eq:zeta}\\
    \sigma_{ij,n}^2 &= \max \qty{\qty|\zeta_{ij,n}|^2, \varepsilon}.\label{eq:epsilon2}
\end{align}
The reason for using the DNN for each channel is that 
each separated signal and each output of the DNN
need to be in phase when we obtain the noise signal by subtraction.

\subsection{Self-Estimation of Prior Distribution of Noise}
We consider the silent parts of the output of Denoiser to be the noise-only parts in the observed signal.
By thresholding the denoised signal by Denoiser, we extract the noise-only signal $\qty{{\bm{x}}_{ij'}}_{j'\in \mathcal{J}'}$, where
\begin{align}
    \mathcal{J}' &= \left\{j\in
    \left\{
        1,\cdots, J
    \right\}
    \middle| \sqrt{\sum_i \abs{\qty(\mathrm{DNN}\qty(X_{n_t}))_{ij}}^2} < \theta
    \right\}.\label{eq:threshold}
\end{align}
Here, $\theta>0$ is a threshold parameter. The SCM of the noise-only time intervals is calculated as
\begin{align}
    \breve{R}_i^{(\mathrm{n})} &= \mathbb{E}\, \qty[{\bm{x}}_{ij'}{\bm{x}}_{ij'}^\mathsf{H}]\nonumber\\
    &\approx \frac{1}{\# \mathcal{J'}}\sum_{j'}{\bm{x}}_{ij'}{\bm{x}}_{ij'}^\mathsf{H},
\end{align}
where $\#$ denotes the cardinality of a finite set.

One of the conjugate prior distributions for the covariance matrix in the multivariate complex Gaussian distribution is the complex inverse matrix gamma distribution (CIMGD) \cite{iranmanesh2013inverted}, which is an extension of the inverse gamma distribution to a positive definite Hermitian matrix. We assume that $\scm{i}{n}$ is generated by the $M$-dimensional CIMGD:
\begin{align}
    &p\qty(\scm{i}{n};\alpha',\beta',\breve{R}_i^{(\mathrm{n})}) \nonumber \\
    &\propto \qty(\det \scm{i}{n})^{-(\alpha' + M)}&\exp(-\frac{1}{\beta'}\tr(\breve{R}_i^{(\mathrm{n})}\qty(\scm{i}{n})^{-1})), \label{eq:cimg}
\end{align}
where $\alpha'>M-1$, $\beta'>0$, and $\breve{R}_i^{(\mathrm{n})}$ are the shape parameter, the scale parameter, and the scale matrix, respectively. If $\alpha'=\tau$ and $\beta'=1$ in (\ref{eq:cimg}), this distribution coincides with the complex inverse Wishart distribution with degree of freedom $\tau$.

\subsection{Log-Posterior with Prior Distribution of Noise}

The log-posterior likelihood given the prior distributions for the variance of target speech and the noise SCM is
\begin{align}
    \mathcal{L}(\Theta) =&\log \prod_{i,j} p(\bm{x}_{ij}|\Theta)p(\covar{ij}{t}; \alpha, \beta)p\qty(\scm{i}{n};\alpha',\beta',\breve{R}_i^{(\mathrm{n})})\nonumber \\
    = &-\sum_{i,j}\Bigg(\bm{x}_{ij}^\mathsf{H}\scm{ij}{o}\bm{x}_{ij} + \log \det \scm{ij}{o} \nonumber\\
    &+ (\alpha + 1)\log \covar{ij}{t} +  \frac{\beta}{\covar{ij}{t}}\Bigg)\nonumber\\ 
    &- \sum _i \Bigg((\alpha'+M)\log \det \scm{i}{n}\nonumber\\
    &+ \frac{1}{\beta'}\tr(\breve{R}_i^{(\mathrm{n})}\qty(\scm{i}{n})^{-1})\Bigg) + \mathrm{const.}
\end{align}
We calculate the $Q$ function in the same manner as in~\cite{kubo2019efficient}.
\begin{align}
    Q\qty(\Theta; \tilde{\Theta}) =& -\sum_{i,j}\Bigg((\alpha + 2)\log \covar{ij}{t} +\frac{\hat{r}_{ij}^{(\mathrm{t})} + \beta}{\covar{ij}{t}}\nonumber \\
    &+M\log \covar{ij}{n}+\log \det \scm{i}{n} \nonumber \\
    & + \frac{\tr(\hat{R}_{ij}^{(\mathrm{n})}\qty(R_{i}^{(\mathrm{n})})^{-1})}{\covar{ij}{n}}\Bigg)\nonumber \\
    & - \sum_i\Bigg(
        \qty(
            \alpha' + M
        )\log \det \scm{i}{n} \nonumber \\
    & +\frac{1}{\beta'}\tr(
            \breve{R}_i^{(\mathrm{n})}\qty(\scm{i}{n})^{-1}
        )
    \Bigg) + \mathrm{const.}\label{eq:proposed Q}
\end{align}
It is easily seen that we only need to change the update rule of $\lambda_i$ in (\ref{eq:lambdaruleold}). Using (\ref{eq:logdetR}) and (\ref{eq:Rinverse}), we focus on the term depending on $\lambda_i$ and modify (\ref{eq:proposed Q}) to
\begin{align}
    Q\qty(\Theta; \tilde{\Theta}) =& \qty(\alpha' + M + J)\log \lambda_i\nonumber \\
    & + \frac{1}{\lambda_i}\bm{u}_i^\mathsf{H}\qty(\frac{1}{\beta'}\breve{R}_i^{(\mathrm{n})} + \sum_j \frac{1}{{r}_{ij}^{(\mathrm{n})}}\hat{R}_{ij}^{(\mathrm{n})})\bm{u}_i\nonumber \\
    &+\mathrm{const.}
\end{align}
We obtain the following update rule, instead of (\ref{eq:lambdaruleold}), by solving ${\partial Q}/{\partial\lambda_i} = 0$:
\begin{align}
    \lambda_i\leftarrow \frac{1}{\alpha'+M+J}\bm{u}_i^\mathsf{H}\qty(\frac{1}{\beta'}\breve{R}_i^{(\mathrm{n})} + \sum_j \frac{1}{\tilde{r}_{ij}^{(\mathrm{n})}}\hat{R}_{ij}^{(\mathrm{n})})\bm{u}_i. \label{eq:lambdarule}
\end{align}

\begin{figure}[t]
\begin{center}
\includegraphics[width=80mm]{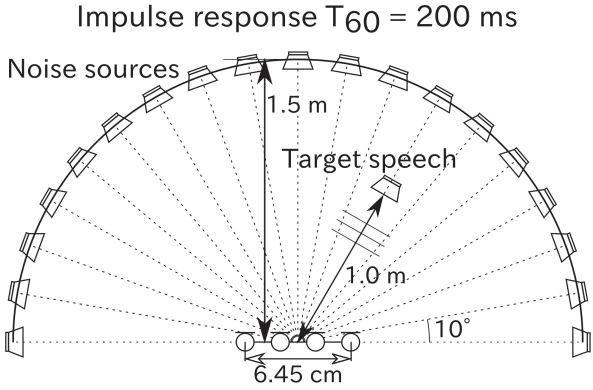}
\end{center}
\vspace*{-13pt}
\caption{Impulse response recording conditions in \cite{kubo2020blind}.}
\label{fig:IRrec}
\end{figure}

\begin{figure*}[ht]
\begin{center}
\includegraphics[width=185mm]{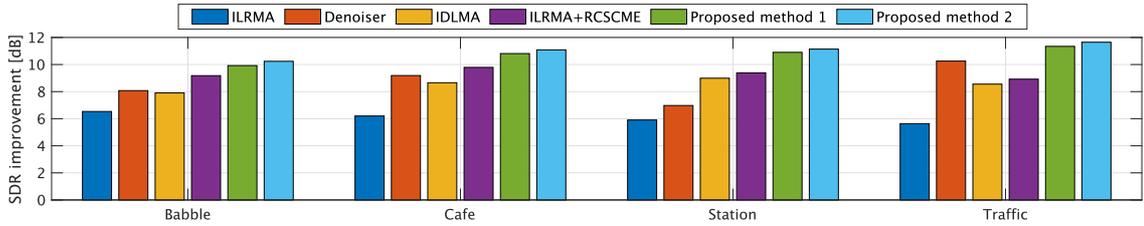}
\end{center}
\vspace*{-13pt}
\caption{Largest SDR improvement among the iterations under each noise.}
\label{fig:sdr vs noise}
\end{figure*}

\begin{figure}[ht]
\begin{center}
\includegraphics[width=85mm]{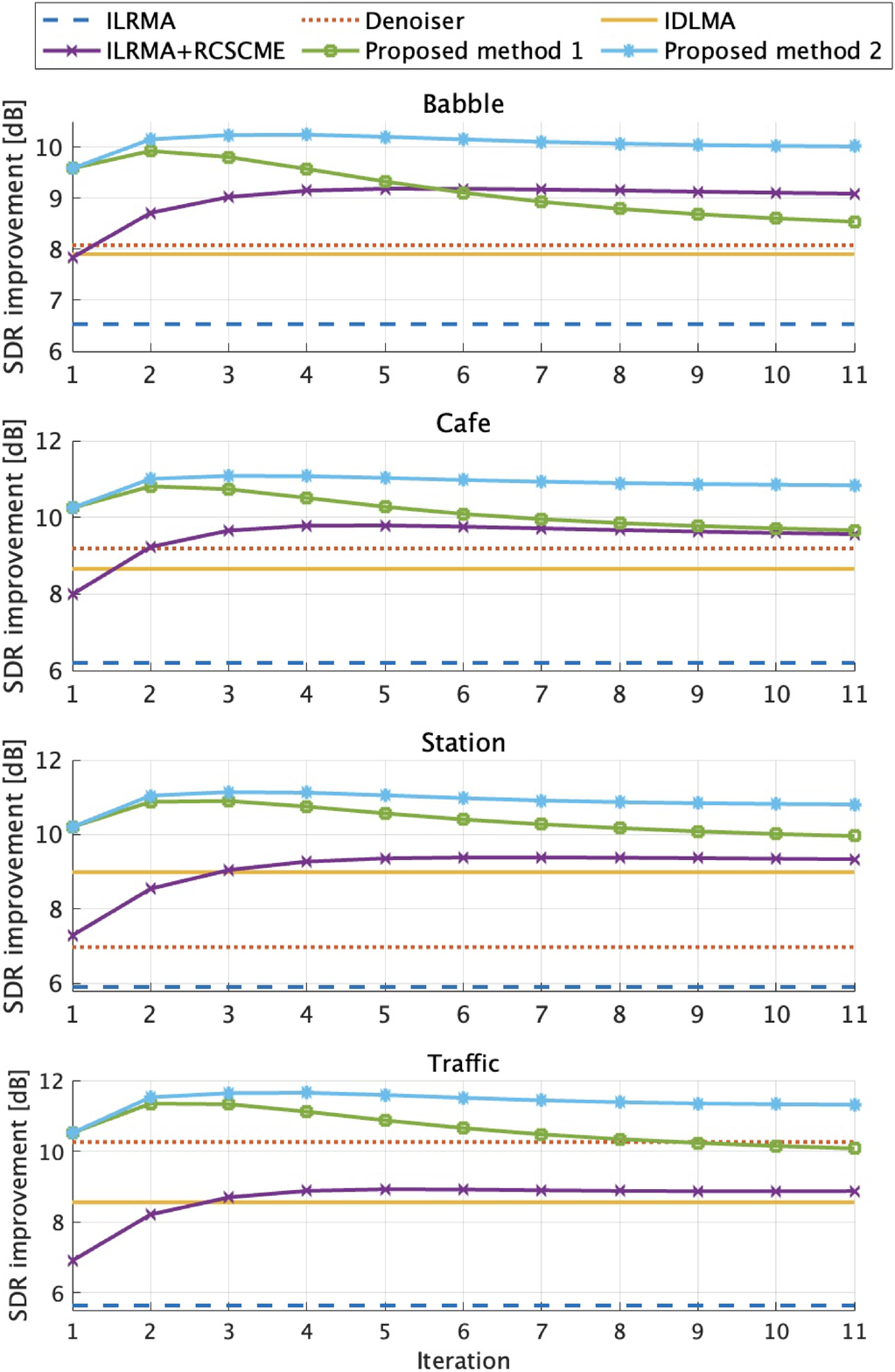}
\end{center}
\vspace*{-15pt}
\caption{Behavior of SDR improvements under each noise.}
\label{fig:sdr vs iteration}
\end{figure}
\section{Experiment}
\subsection{Experimental Conditions}
We conducted a simulated experiment to verify the effectiveness of the proposed method. 
The diffuse noise was simulated by convolution of the impulse response from 19 loudspeakers to four microphones. 
We used the four noise signals: babble, cafe, station, and traffic noise.
For the babble noise, we used 19 speech signals from the JNAS speech corpus \cite{itou1999jnas}.
For the cafe, station, and traffic noise, we used 19 signals, which were obtained by dividing each noise signal from DEMAND \cite{thiemann2013demand}.
The target speech signal was simulated by convolution of the impulse response from a closer location. 
For the target speech signal, we used one speech signal from the JNAS speech corpus \cite{itou1999jnas}.
We used the conditions in Fig.~\ref{fig:IRrec} to record the impulse response~\cite{kubo2020blind}.
The sampling rate was 16~kHz. 
The input signal-to-noise ratio was set to 0~dB.
The STFT was performed by using a 64-ms-long Hamming window and 32-ms-long shift.

We compared ILRMA, IDLMA, RCSCME using ILRMA (ILRMA+RCSCME), RCSCME using IDLMA with $\sigma_{ij,n}$ updated by (\ref{eq:zeta}) and (\ref{eq:epsilon2}) (proposed method 1), and noise self-supervised RCSCME using IDLMA with $\lambda_i$ updated by (\ref{eq:lambdarule}) (proposed method 2). 
We used source-to-distortion ratio (SDR) \cite{vincent2006performance} improvement as a measure. We used $(\alpha, \beta, \alpha', \beta')=(1.3, 10^{-16}, 8\times 10^2, 10^4)$ for each method, which were selected experimentally. The flooring parameter in (\ref{eq:epsilon1}) and (\ref{eq:epsilon2}) was set to $\varepsilon = 0.1  \times \sum_{i,j}\sigma_{ij,n}^2/(IJ)$ and the threshold parameter in (\ref{eq:threshold}) was set to $\theta = 10^{-3}$. The demixing matrix $W_i$ in ILRMA and IDLMA was initialized by the identity matrix. 
The number of NMF bases of ILRMA was set to $10$ and the NMF variables were initialized by random values taken from the uniform distribution on $[0,1]$.
We used the unit eigenvector corresponding to the zero eigenvalue of $\scmd{i}{n}$ as $\bm{v}_i$ in RCSCME. For Denoiser, we used the trained model provided by \cite{defossez2020real} with $H=48$ hidden channels. Since Denoiser is the waveform-domain model, we used Denoiser as
\begin{align}
    \mathrm{DNN}(Y_n) = \mathrm{STFT}\qty[\mathrm{Denoiser}\qty(\mathrm{STFT}^{-1}\qty[Y_n])].
\end{align}
We set the number of iterations in ILRMA to 50, that in IDLMA to 90, and that in RCSCME for ILRMA+RCSCME, proposed method 1, and proposed method 2 to 10. 
In IDLMA, $\sigma_{ij,n}$ was updated by DNN for every 30 times $W_i$ was updated.
These numbers of iterations were experimentally determined.

\subsection{Results}
We took the average of the SDR improvement results for 10 different sets of randomly initialized values
of the NMF variables in ILRMA and ILRMA+RCSCME.
The SDR improvements for each noise are shown in Fig. \ref{fig:sdr vs noise}.
For each RCSCME method, we show the largest SDR improvement among the iterations. 
As shown in Fig \ref{fig:sdr vs noise}, Denoiser outperformed IDLMA for babble, cafe, and traffic noise. The reason is that 
Denoiser includes a nonlinear transformation that can provide stronger noise reduction than linear systems, whereas IDLMA
estimates the output signals via a linear time-invariant spatial filter, i.e., demixing matrix $W_i$. 
However, proposed method 1 shows better performance than Denoiser and IDLMA for all the noises, implying the advantage of the time-variant properties in RCSCME.
It is also shown that proposed method 2 achieves the best separation performance among all methods; this would be due to introduction of the self-estimated prior distribution of the noise SCM.
The SDR improvement for each iteration is shown in Fig. \ref{fig:sdr vs iteration}. 
Fig. \ref{fig:sdr vs iteration} shows that ILRMA+RCSCME requires more than five iterations to converge, while proposed method 1 and proposed method 2 take the maximum value in less than five iterations.

\section{Conclusions}
We proposed RCSCME using IDLMA, which is applied to speech enhancement by preparing a single-channel speech enhancement DNN for estimating the power spectrogram of the target speech and the noise.
We used Denoiser as the single-channel speech enhancement DNN model for estimating the power spectrogram of the target speech. To estimate the power spectrogram of diffuse noise, we subtracted the speech extracted by Denoiser from the separated signal.
We also proposed a method to estimate the noise-only time intervals using the output of Denoiser and design the prior distribution of the noise SCM for RCSCME. 
We confirmed that noise self-supervised RCSCME using IDLMA outperformed the other methods and took the maximum value in fewer iterations than the conventional methods.

\section*{Acknowledgment}
This work was supported by the Japan--New Zealand Research Cooperative Program of JSPS and RSNZ (Grant Number JPJSBP120201002), JSPS KAKENHI Grant Numbers 19K20306, 19H01116, and 19H04131, and JST Moonshot R\&D Grant Number JPMJPS2011.

\bibliographystyle{IEEE}
\bibliography{main}
\end{document}